\documentclass[iop]{emulateapj}
\usepackage{epsfig}
\usepackage{epstopdf}
\usepackage{natbib}
\usepackage{amssymb}
\usepackage{amsbsy}
\usepackage{subfigure}
\usepackage[mathcal]{euscript}
\usepackage{float}
\usepackage{amsmath}
\usepackage{tabularx}
\usepackage{xspace}
\usepackage{enumitem}
\usepackage{stackengine}
\usepackage{comment}
\usepackage[dvipsnames]{xcolor}

\usepackage{ulem,xspace}

\newcommand{\msun}{M$_\sun$}

\newcommand{\kepler}{\textit{Kepler}}
\newcommand{\tess}{\textit{TESS}}
\newcommand{\gaia}{\textit{Gaia}}

\newcommand{\teff}{\mbox{$T_{\rm eff}$}}
\newcommand{\logg}{\mbox{$\log g$}}
\newcommand{\feh}{\mbox{$\rm{[Fe/H]}$}}
\newcommand{\fbol}{\mbox{$f_{\rm bol}$}}

\newcommand{\kiauhoku}{\texttt{kiauhoku}} 

\bibpunct{(}{)}{;}{a}{}{,}

\begin{document}

\title{A Guide to Realistic Uncertainties on Fundamental Properties of Solar-Type Exoplanet Host Stars}

\author{
Jamie Tayar\altaffilmark{1,2},
Zachary R. Claytor\altaffilmark{1},
Daniel Huber\altaffilmark{1},
Jennifer van Saders\altaffilmark{1}
}
\altaffiltext{1}{Institute for Astronomy, University of Hawai`i, 2680 Woodlawn Drive, Honolulu, Hawaii 96822, USA}
\altaffiltext{2}{Hubble Fellow}

\begin{abstract}
Our understanding of the properties and demographics of exoplanets critically relies on our ability to determine fundamental properties of their host stars. The advent of \gaia\ and large spectroscopic surveys has now made it in principle possible to infer properties of individual stars, including most exoplanet hosts, to very high precision. However, we show that in practice, such analyses are limited both by uncertainties in the fundamental scale, and by uncertainties in our models of stellar evolution, even for stars similar to the Sun. For example, we show that current uncertainties on measured interferometric angular diameters and bolometric fluxes set a systematic uncertainty floor of $\sim$2\% in temperature, $\sim$2\% in luminosity, and $\sim$4\% in radius. Comparisons between widely available model grids suggest uncertainties of order $\sim$\,5\% in mass and $\sim$\,20\% in age for main sequence and subgiant stars. While the radius uncertainties are roughly constant over this range of stars, the model dependent uncertainties are a complex function of luminosity, temperature, and metallicity. We provide open-source software for approximating these uncertainties for individual targets, and discuss 
strategies for reducing these uncertainties in the future.
\end{abstract}


\section{Introduction}
\setcounter{footnote}{0}

Answering questions about the formation, evolution, composition, and habitability of exoplanets requires large samples of precise measurements of planetary properties such as mass, radius, and age. Since most of these properties are measured relative to their host stars, such work also requires detailed stellar characterization. For example, the irradiation-dependent planet radius gap between super-Earth and sub-Neptune sized planets was only recently quantified because the feature is narrow enough to only be visible in samples with careful spectroscopic \citep{Fulton2017}, asteroseismic \citep{vaneylen18}, or astrometric \citep{berger18,ullman20} characterization. Even so, the dominant mechanism causing this gap is still debated, with core-powered mass loss \citep{ginzburg18} and photoevaporation \citep{owen17} both making predictions consistent with the observations that can only be distinguished by even more precise samples.

More generally, the availability of high-resolution spectroscopy and high-precision \gaia\ parallaxes  has recently allowed claims of extremely precise exoplanet host star properties, with uncertainties on mass and radius approaching and sometimes reaching below one percent. While such small uncertainties would be a boon to exoplanet demographic studies, they raise questions about whether the fundamental systematic uncertainties have been fully considered. In general, neither  stellar mass nor radius can be measured directly, which suggests that there is likely to be a floor in how precisely a star's mass and radius can be estimated. 

Stellar radii, for example, are often inferred from a combination of parallaxes and either photometric or spectroscopic estimates of temperature and metallicity. Such estimates rely on bolometric corrections, reddening maps, and stellar atmosphere models, all of which have been shown to be uncertain \citep{torres12,GonzalezHernandezBonifacio2009, casagrande20}. On top of that, uncertainty in the fundamental temperature scale computed from stars with measured angular diameters adds additional complexity.


Stellar mass estimates are often even more indirect, with stellar models being used to infer mass from the inferred 
luminosity, effective temperature, and composition. 
Numerous model grids are publicly available to allow this (e.g. MIST, \citealt{Choi2016}, PARSEC, \citealt{Bressan2012}, DSEP, \citealt{Dotter2008}, Yonsai\textemdash Yale, \citealt{Spada2013}, etc.) and a variety of tools have been developed to simplify the inference such as \texttt{isochrones} \citep{Morton2015}, \texttt{isoclassify} \citep{Huber2017isoclassify,berger20a}, PARAM \citep{dasilva06,rodrigues14}, and \texttt{exofast} \citep{eastman13,Eastman2019}. 

Previous work has shown that the choice of stellar modeling code has at most a very small effect on model predictions, as long as the exact same physics is used \citep{SilvaAguirre2020}. However, the generation of these grids of stellar models requires many physical choices to be made. Convection, for example, is an inherently three-dimensional process that must be parameterized into one dimension, and different model grids have made different choices of how to do this \citep{Tayar2017}. 
Similarly, choices about the atmosphere boundary condition \citep{Choi2018a}, composition \citep{vanSadersPinsonneault2012, CapeloLopes2020}, rotation \citep{vanSadersPinsonneault2013},  opacities \citep{Valle2013a}, overshoot \citep{Pedersen2018}, and so on can impact the models.

Since much of this physics is uncertain at a significant level, modelers often make perfectly reasonable but slightly different physical choices that result in offsets between different grids of stellar models. In the past, such offsets were substantially less important than the observational uncertainties. However, with the ability to calculate extremely precise luminosities using \gaia\ Data Release 2 \citep{Lindegren2018} combined with the availability of high-resolution, high signal-to-noise spectra for estimating temperatures and metallicities, the observational uncertainties on estimated stellar parameters can now be comparable to or smaller than the systematic uncertainties coming from model physics or the fundamental temperature scale. 
Incorporating such systematic host star uncertainties is critical to derive reliable uncertainties on fundamental properties of exoplanets, especially when samples from different studies are combined \citep[for example through the NASA Exoplanet Archive,][]{akeson13} to study exoplanet demographics, or when interpreting observations of exoplanet atmospheres.

In this paper, we aim to provide a guide to realistic uncertainties on fundamental properties of exoplanet host stars by investigating sources of systematic errors both on observable quantities (such as temperature, radius and luminosity) and those that are estimated from evolutionary models (such as mass and age). We note that there are additional complexities in low-mass stars \citep{kraus11}, massive stars \citep{Holgado2020}, and pre-main-sequence stars \citep{SomersStassun2017}, that make them more challenging to characterize and thus their error floor is likely higher. We therefore focus only on solar-type (FGK) stars here, which in theory should be easier to work with, but practically still have significant uncertainty. 

\section{Uncertainties from Input Observables}


\subsection{Bolometric Fluxes and Luminosities}

The most readily available observations for a given exoplanet host star are broadband photometry in optical and near-infrared wavelengths such as Tycho-2 \citep{hog00}, 2MASS \citep{skrutskie06} and \gaia\ \citep{evans18} and a high-precision parallax from \gaia\ \citep{Lindegren2018}. The former is typically used to estimate the bolometric flux received on Earth ($f_{\rm bol}$), either by approximating the spectral energy distribution (SED) by integrating fluxes from broadband photometry in combination with model atmospheres \citep[SED fitting, e.g.][]{vanbelle09,huber12,mann13,stassun16a} or by applying bolometric corrections derived from model atmospheres \citep[e.g.][]{alonso04,torres10,casagrande18}. 
The latter requires an estimate of \teff\, which is typically obtained through calibrated color-\teff\ relations \citep{casagrande11,casagrande20}, while SED fitting typically simultaneously solves for \teff\ and \fbol\ or uses external \teff\ estimates from high-resolution spectroscopy. The combination of \fbol\ with the distance $d$ estimated from the parallax \citep[e.g.][]{bailer15} then allows a measurement of the luminosity:

\begin{equation}
L = 4 \pi d^{2} f_{\rm bol} \: .
\end{equation}

Typical fractional distance uncertainties for exoplanet host stars in the \gaia\ era are $\ll$\,1\,\%, and therefore negligible. The error floor for \fbol\ is set by the accuracy of photometric zeropoints, which are typically known to 1-2\% for ground-based photometry \citep{mann15,casagrande18,gaiaphot} based on comparisons with space-based spectrophotometry from HST/STIS \citep{bohlin14}. Ground-based photometry obtained from different surveys can also exhibit substantial offsets at the $\gtrsim$\,0.01\,mag level, as shown for example in the various photometric surveys which have targeted the \kepler\ field \citep[e.g.][]{pinsonneault11,greiss12}. Therefore, care should be taken when combining literature photometry without consideration of the accuracy of input photometry. Another source of uncertainty are differences in predicted fluxes between stellar model atmospheres, which can reach up to $\sim$\,5\% \citep[e.g. Appendix A in][]{zinn19}.  

\begin{figure}
\begin{center}
\resizebox{\hsize}{!}{\includegraphics{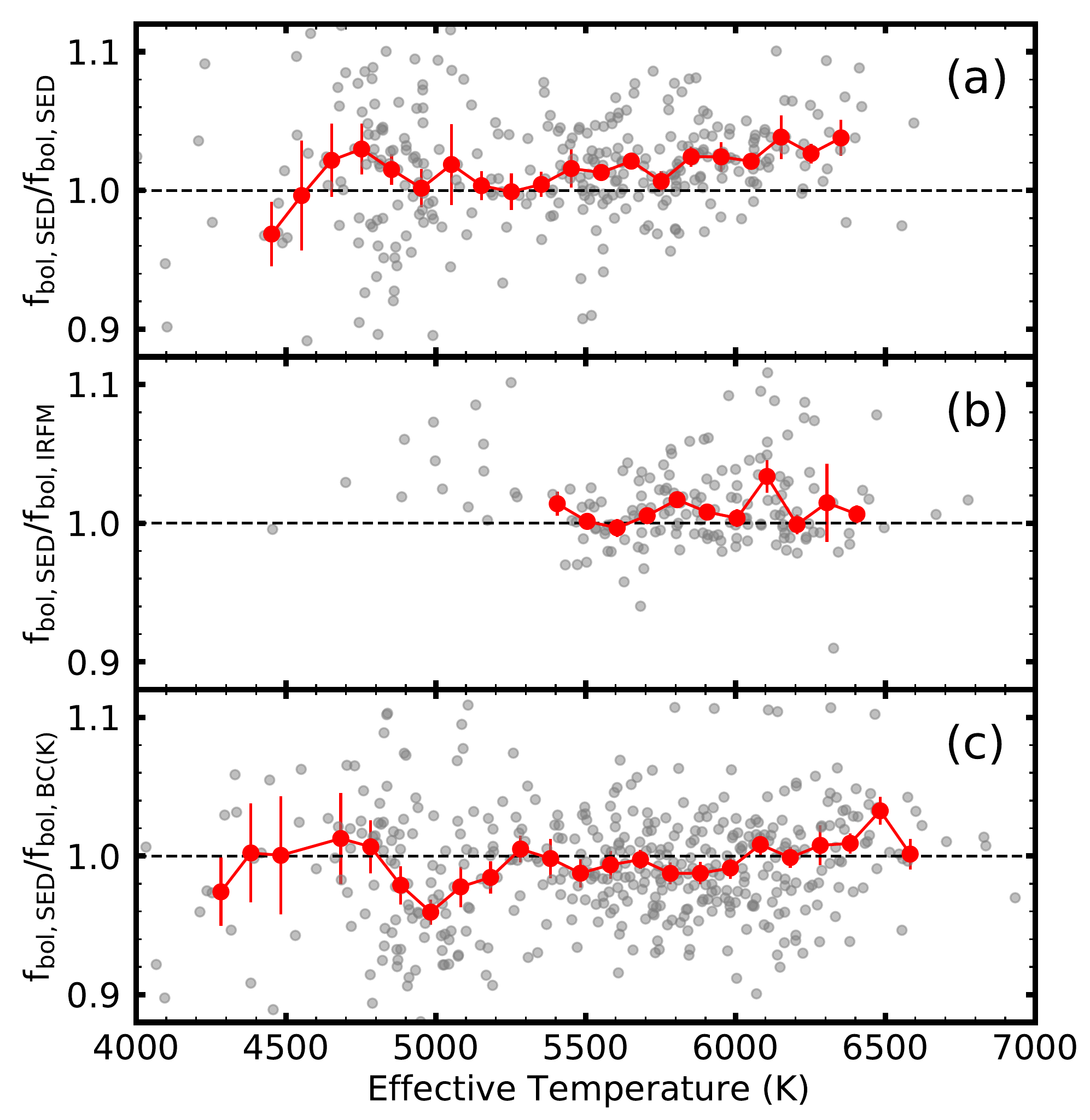}}
\caption{Bolometric fluxes for exoplanet host stars derived using SED fitting from \citet{stassun17} compared to an alternative SED-fitting methodology \citep[panel a,][]{mcdonald17}, the infrared flux method \citep[panel b,][]{casagrande11} and $K$-band bolometric corrections from the MIST grid \citep{Choi2016} as implemented in \texttt{isoclassify} \citep[panel c,][]{huber17}. Grey points show individual stars, and red circles show binned averages in steps of 100\,K. The typical random scatter and systematic offsets between methods as a function of \teff\ are between $2-4$\%, setting a fundamental limit on the uncertainty of \gaia-derived luminosities.}
\label{fig:bolflux}
\end{center}
\end{figure}

Bolometric flux measurements require corrections for interstellar reddening, which can be measured if supplementary information on \teff, \logg\ and \feh\ are available from spectroscopy and/or asteroseismology to constrain the shape of the SED \citep{rodrigues14, huber17}. If only broadband photometry and parallaxes are available, reddening and \teff\ are degenerate and unphysical extinction values may result from compensating for differences between models and observations which are not actually related to extinction. This can be partially avoided through the use of infrared photometry and 3D extinction maps \citep[e.g.][]{green16}, although the latter may suffer from significant systematic errors, particularly for nearby stars (D. Godoy-Rivera et al., submitted). In summary, typical uncertainties due to reddening corrections are at the $\sim$\,0.02\,mag level, comparable to photometric zeropoint offsets \citep{huber17}.

Figure 1 illustrates the combined effects of these uncertainties by comparing \fbol\ estimates from SED fits for exoplanet host stars from \citet{stassun17}, which are commonly adopted as default stellar properties on the NASA Exoplanet Archive, to independent \fbol\ measurements using SED fitting \citep[Fig.\ 1a,][]{mcdonald17}, the infrared flux method \citep[Fig.\ 1b,][]{casagrande11} and interpolating $K$-band bolometric corrections from the MIST grid \citep{Choi2016} as implemented in \texttt{isoclassify} \citep[Fig.\ 1c,][]{huber17, berger20a}. The typical scatter ranges between $2-4$\%, with systematic offsets as a function of effective temperature reaching up to $\sim$\,4\%. Note that \fbol\ in Figure 1c was derived using the same \teff\ and extinction values, and thus solely reflects systematic differences between photometric zeropoints and model atmospheres. On average, this comparison demonstrates that contributions of photometric zeropoints, model atmosphere grids, and reddening sets a fundamental floor of $\sim\,2\%$ on bolometric fluxes (and thus luminosities) for a given exoplanet host star. 


\subsection{Angular Diameters and Effective Temperatures}

The effective temperature of a star is defined through its bolometric flux and angular diameter $\theta$:

\begin{equation}
T_{\rm eff} = \left(\frac{4 f_{\rm bol}}{\sigma \theta^2}\right)^{1/4} \: .
\end{equation}

A fundamental \teff\ measurement thus requires measurements of both \fbol\ and the angular diameter. \textit{Temperatures from high-resolution spectroscopy, color-\teff\ relations or SED fitting have to be calibrated using stars with measured angular diameters.} The internal consistency of measured angular diameters then sets a fundamental limit as to how well \teff\ (and in combination with luminosity, radius) can be determined.

The most successful method to resolve the small angular sizes of stars is optical long-baseline interferometry using facilities such as the Center for High Angular Resolution (CHARA) Array \citep{brummelaar04}, the Navy Precision Optical Interferometer \citep[NPOI,][]{Armstrong2013JAI.....240002A} and the Very Large Telescope Interferometer (VLTI). A few hundred stars have measured diameters with uncertainties of a few percent \citep{bb17}. The accuracy of angular diameter measurements relies on calibration to account for the temporal and spatial reduction of fringe visibilities due to atmospheric turbulence, with major sources of systematic errors including the estimates of calibrator diameters, under-resolving target stars, and uncertainties in the adopted wavelength scale \citep[e.g.][]{vanbelle05}. Recent diameter measurements using different instruments have shown significant systematic offsets \citep{white18,karovicova18}, which are important since many indirect methods are calibrated on a small number of stars with published angular diameters \citep{casagrande14c, casagrande20}. An example of this ``calibration pyramid'' is the infrared color-\teff\ relation by \citet{GonzalezHernandezBonifacio2009}, which is used to calibrate temperatures for hundreds of thousands of giants in large spectroscopic surveys such as APOGEE \citep{Majewski2017}, but hinges on a handful of measured angular diameters with unknown systematic errors.

\begin{figure}
\begin{center}
\resizebox{\hsize}{!}{\includegraphics{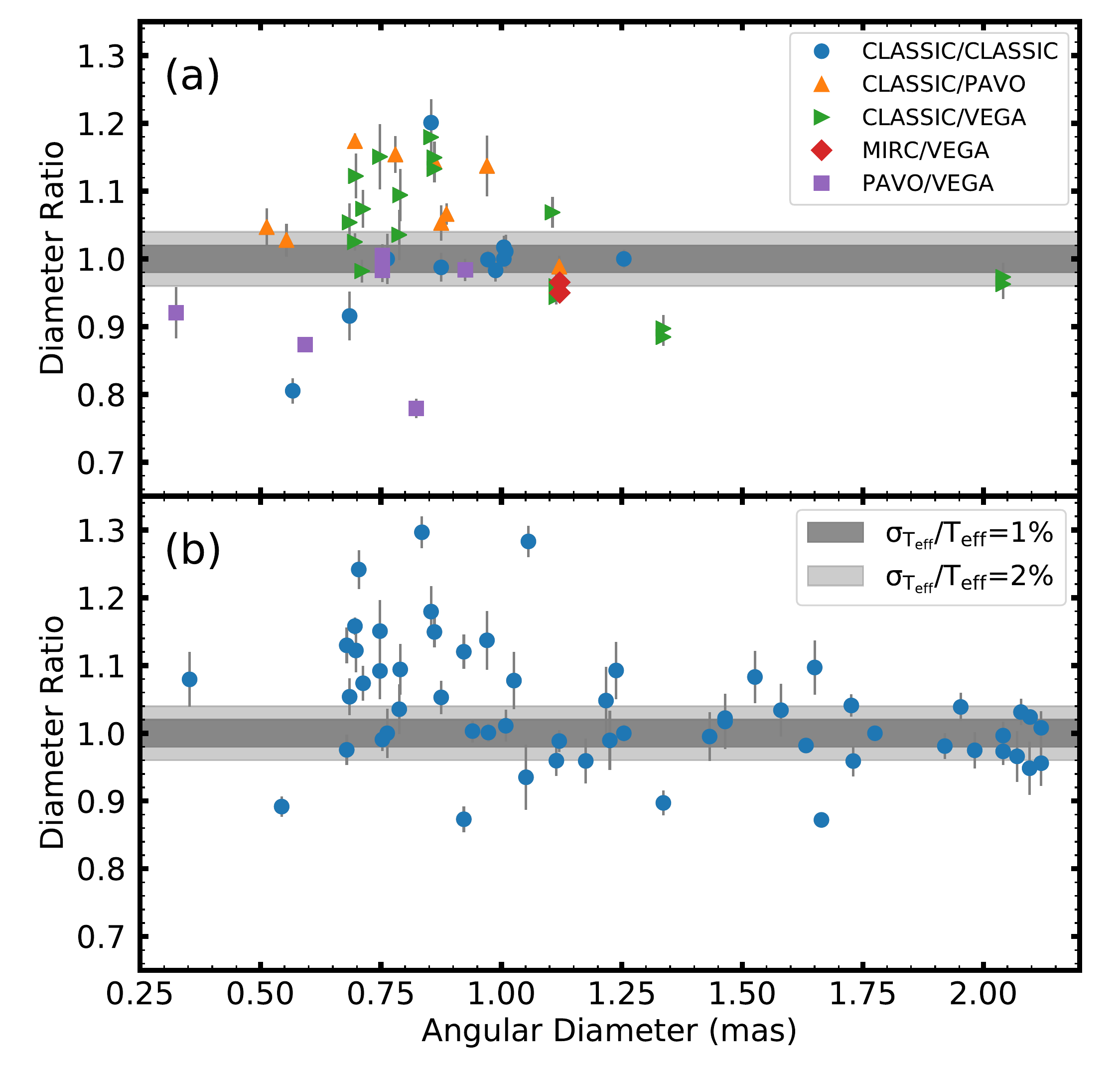}}
\caption{Panel (a): Ratio of interferometric angular diameters from the CHARA Array for stars with multiple published measurements as a function of diameter. Colors and symbols compare different beam combiners used to obtain the measurements (see text for details). Only measurements with formal uncertainties $<$\,5\% are shown. Panel (b): same as panel (a) but for angular diameters listed in the JMMC Measured Stellar Diameters Catalogue \citep{jmdc}.}
\label{fig:angdia}
\end{center}
\end{figure}

Figure \ref{fig:angdia}a compares angular diameter measurements from the CHARA array for stars with multiple published results in the literature \citep{akeson09, baines08, baines09,baines10, bazot11, berger06, boyajian08, boyajian12, boyajian12b, boyajian13, challouf14, creevey12, creevey15, howard14, johnson14, kane15b, karovicova18, ligi12, ligi16, maestro13, vonbraun11, vonbraun14, white13, white18}. The comparison is grouped by the beam combiners used to obtain the measurements: CLASSIC \citep{brummelaar04}, MIRC \citep{mirc}, VEGA \citep{mourard06} and PAVO \citep{ireland08}. We observe individual systematic differences of up to $\sim 10\%$, which correlate with the instrument combination used for the measurement. Figure \ref{fig:angdia}b compares a larger sample of measurements from the JMMC Measured Stellar Diameters Catalogue \citep{jmdc}, showing a similar result and a trend of increasing systematic errors with decreasing angular size. The latter implies that the dominant source of systematic error is under-resolving target stars, which for a given angular size is more severe for infrared than optical beam combiners since infrared wavelengths yield lower angular resolution at a given baseline. While recent comparisons from CHARA and VLTI have shown more promising agreement \citep[][Creevey et al., in prep]{rains20}, it is clear that a larger number of diameter measurements of the same stars with different instruments are needed to pin down sources of systematic errors in literature measurements, and that \teff\ calibrations require careful sample selection \citep{casagrande14c}. We note that the vast majority of main-sequence stars have angular sizes $<$1\,mas, and thus are most affected by the biases described above.

The systematic differences in Figure 2 sets a fundamental limit on the accuracy of effective temperature scales and thus stellar radii derived from \gaia\ parallaxes. To illustrate this, the dark grey band in Figure \ref{fig:angdia} shows the required accuracy to reach a 1\% calibration in \teff, which is smaller than the systematic differences in the measurements. The median absolute systematic offset over all instrument combinations in Figure \ref{fig:angdia}a is $\sim$\,4\%, which corresponds to an uncertainty in \teff\ of $\sim$\,2\%, a factor of $\sim$\,2-4 higher than typical \teff\ uncertainties quoted for exoplanet host stars in the literature. 
The current sample of interferometric angular diameters thus sets a fundamental limit of $\sim$\,2\% on the effective temperature scale for solar-type stars ($\approx$\,110K at solar \teff), which does not include systematic uncertainties in common observational measurement techniques \citep[e.g][]{Spina2020}.


\subsection{Summary Recommendation}

The comparisons discussed in this section\footnote{All data and code to reproduce the results are available at \url{https://github.com/danxhuber/hoststaruncertainties}} demonstrate that the typical limits for measurements of bolometric fluxes and angular diameters, set by uncertainties in photometric zeropoints, model atmospheres, extinction and interferometric calibration, are currently $\sim$\,2\% and $\sim$\,4\%, respectively. This directly sets lower limits on the uncertainty of ``observed'' fundamental properties of stellar luminosity ($\sim$\,2\%) and effective temperature ($\sim$\,2\%), and thus stellar radius ($\sim$\,4\%). We recommend that these uncertainties are added in quadrature to the formal uncertainties for exoplanet host stars to account for methodology-specific differences, unless uncertainties have already been estimated from multiple independent methods, properties have been directly measured from space-based spectrophotometry or long-baseline interferometry, or for solar twins where measurements have been obtained differentially with respect to the Sun. We note that spectroscopic abundances, which are also used as input observables for evolutionary models and not discussed here explicitly, show similar method-specific differences that are typically larger than formal uncertainties \citep[e.g.][]{torres12,hinkel16}.

\section{Uncertainties from Model Grids}
Estimates of stellar masses are often even less fundamental. In many cases, a combination of luminosity, temperature, and metallicity are compared to a grid of stellar models, and the best match is used to read off the likely mass and age of the star in question. However, we contend that the answer returned depends on the physical assumptions used to construct the stellar evolution model. 


\subsection{Model Physics}
In order to estimate the theoretical uncertainties on estimates of stellar masses and ages as a function of luminosity, effective temperature, and composition, we compare the predictions of several widely used model grids. For each grid, we use models between 0.6 and 2.0 \msun\ at intervals of 0.1 \msun. We also run the analysis at [Fe/H]= -1.0, -0.5, 0.0, and +0.5 as defined by each model. We list the different model grids used in this work and summarize the physics in Table \ref{Table:physics}. We recognize that there are many other grids of models available, but we believe that the grids considered are a representative sample of the choices of model physics and calibration commonly used for the characterization of solar-type stars. We note that this comparison excludes pre-main-sequence stars as well as M dwarfs, as both of these regimes have results that can be even more dependent on the assumed model physics, and models can be significantly discrepant with the observed stellar properties. 



\begin{table*}[htbp]
\begin{minipage}{1.0\textwidth}
\caption{Summary of the input physics used in each model.} 
\begin{tabularx} {.98\textwidth}{>{\raggedright\arraybackslash}X|>{\raggedright\arraybackslash}X|>{\raggedright\arraybackslash}X|>{\raggedright\arraybackslash}X|>{\raggedright\arraybackslash}X} 
\hline\hline
Parameter & YREC & MIST & DSEP & GARSTEC \\ \hline
Reference & This work & \citet{Choi2016} & \citet{Dotter2008} & \citet{Serenelli2013}\\
Atmosphere & grey & \citet{Kurucz1993}& PHOENIX \citep{Hauschildt1999a,Hauschildt1999b}& grey\\ 
Convective Overshoot & Step: 0.16H$_{\rm p}$ & Diffusive: 0.0160(core) and 0.0174(env) & Step: 0.2H$_{\rm p}$ & Diffusive: 0.02\\ 
Diffusion & Yes & Main Sequence only& Modified & Yes\\ 
Equation of State & OPAL+SCVH & OPAL+SCVH+ MacDonald+HELM+PC & Ideal Gas with \citet{ChaboyerKim1995}+ \citet{Irwin2004}& \citet{Irwin2004}\\ 
High-Temperature Opacities& OPAL  & OPAL & OPAL& OPAL\\ 
Low-Temperature Opacities & \citet{Ferguson2005} & \citet{Ferguson2005}& \citet{Ferguson2005} & \citet{Ferguson2005}\\ 
Mixing Length & \citet{Tayar2017} & 1.82 & 1.938 & 1.811\\ 
Mixture and Solar Z/X & \citet{GrevesseSauval1998} & \citet{Asplund2009} protosolar & \citet{GrevesseSauval1998} & \citet{GrevesseSauval1998}\\
Nuclear Reaction Rates & \citet{Adelberger2011} & \citet{Cyburt2010} & \citet{Adelberger1998}+\citet{Imbriani2004}+\citet{Kunz2002}+\citet{Angulo1999} & \citet{Adelberger1998}+\citet{Angulo1999}\\ 
Rotation & \citet{TayarPinsonneault2018} & None & None & None \\
Weak Screening & \citet{Salpeter1954} & \citet{AlastueyJancovici1978} & \citet{Salpeter1954}+ \citet{Graboske1973} & \citet{Salpeter1954}\\ 
Solar X & 0.709452 & 0.7154 & 0.7071& 0.7090 \\ 
Solar Y & 0.2725693 & 0.2703 & 0.27402 & 0.2716 \\ 
Solar Z & 0.0179492 & 0.0142 & 0.01885& 0.0193\\ 
$\Delta$Y/$\Delta$Z & 1.3426 & 1.5 & 1.5327 & 1.194 \\ 
Surface (Z/X)$_\sun$ & 0.0253 & 0.0173 & 0.0229 & 0.0245 \\
\hline
\end{tabularx}
\label{Table:physics}
\end{minipage}
\end{table*}
\subsubsection{YREC Models}
The Yale Rotating Evolution Code \citep[YREC,][]{Pinsonneault1989} grid of models are the only grid of models used in the work presented here for the first time, and represent an expansion of the grid presented in \citep{TayarPinsonneault2018}. These models are unique in that they are not calibrated to the sun, but instead designed to replicate the observed properties of stars on the red giant branch as a function of metallicity \citep{Tayar2017}. 
These models use a \citet{GrevesseSauval1998} chemical mixture with a helium enrichment of $\frac{\Delta Y}{\Delta Z}$=1.3426 \citep{Tayar2017}. They have a convective step overshoot of 0.16 H$_{\rm p}$, calibrated on the luminosity of the secondary red clump \citep{TayarPinsonneault2018}, and rotational evolution that includes angular momentum loss according to the Pinsonneault, Matt, and Macgregor wind loss law \citep{vanSadersPinsonneault2013} as described in \citet{TayarPinsonneault2018} for stars below the Kraft break, but they do not include rotational mixing. The models do include diffusion \citep{BahcallLoeb1990} as implemented for \citet{SomersPinsonneault2016}, use a \citet{Kurucz1997} atmosphere boundary condition and rely on OPAL \citep{IglesiasRogers1996} high-temperature opacities and \citet{Ferguson2005} low-temperature opacities. 
They also use the SCVH \citep{Saumon1995} and OPAL \citep{RogersNayfonov2002} equations of state.

\subsubsection{MIST Models}
The MESA Isochrones and Stellar Tracks \citep[MIST,][]{Choi2016} models are used here in the form of models rather than isochrones. Their creation and properties are described extensively in \citet{Choi2016}. In brief, they are generated using the MESA stellar evolution code \citep{Paxton2011,Paxton2013, Paxton2015, Paxton2018, Paxton2019} and are available online\footnote{http://waps.cfa.harvard.edu/MIST/model\_grids.html}. The solar calibration for these models was chosen to minimize the combined offsets from solar values in log$L$, log$R$, surface composition, the base of the surface convection zone, and the sound speed at 4.57 Gyrs. The composition is based on \citet{Asplund2009}, and diffusion of helium and heavy elements is included for main sequence stars following the \citet{Thoul1994} formalism, balanced by radiation turbulence \citep{MorelThevenin2002}. Exponential overshoot is included for both convective cores and convective envelopes. The non-rotating version of the model grid was used for this analysis, chosen because most of the low-mass stars studied here are expected to be relatively slowly rotating, but a rotating grid is also available, and work is ongoing to update these grids, including the addition of more sophisticated rotation physics \citep{Gossage2020}.

\subsubsection{DSEP}
The Dartmouth Stellar Evolution Program (DSEP) models are used here as presented in \citet{Dotter2008} and available online\footnote{http://stellar.dartmouth.edu/models/grid\_old.html}. In these models, atomic diffusion and gravitational settling are included, but are inhibited in the outer 0.01\msun\ as described in \citet{Chaboyer2001}. The boundary condition for these models is based on a grid of PHOENIX model atmospheres \citep{Hauschildt1999a, Hauschildt1999b} matched at the point where T=T$_{\rm eff}$. 
While models are available for a range of [$\alpha$/Fe] values, we use only [$\alpha$/Fe]=0 models in this work. 

\subsubsection{GARSTEC Models}
The grid of models made using the Garching Stellar Evolution Code \citep[GARSTEC,][]{WeissSchlattl2008} used here was first presented in \citet{Serenelli2013}. The mixing length and reference composition were chosen to match the parameters of a solar model at the solar age, including the effects of diffusion. Convective overshooting is modeled diffusively following the prescriptions of \citet{Freytag1996}, although limits are placed to prevent extensive overshooting in small convective cores \citep{Magic2010}. Updated versions of this grid have been used for Bayesian estimates of stellar parameters including asteroseismic inputs \citep[e.g.][]{Serenelli2017}, but we use them here with only classical parameters.


\begin{figure}[ht]
\begin{center}
\subfigure{\includegraphics[width=0.45\textwidth,clip=true, trim=0.5in 0in 0in 0in]{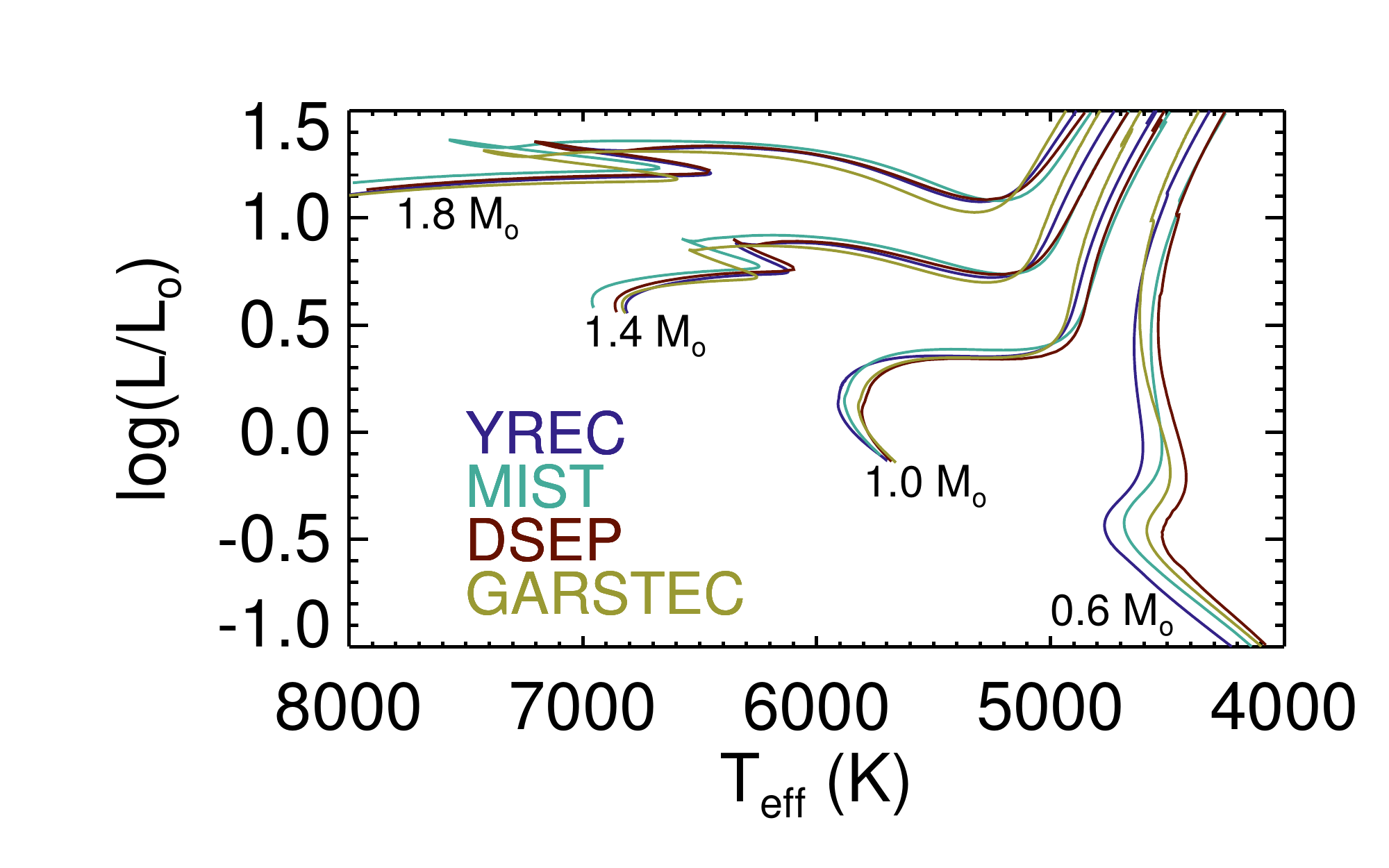}}
\subfigure{\includegraphics[width=0.45\textwidth,clip=true, trim=0.35in 0in 0in 0in]{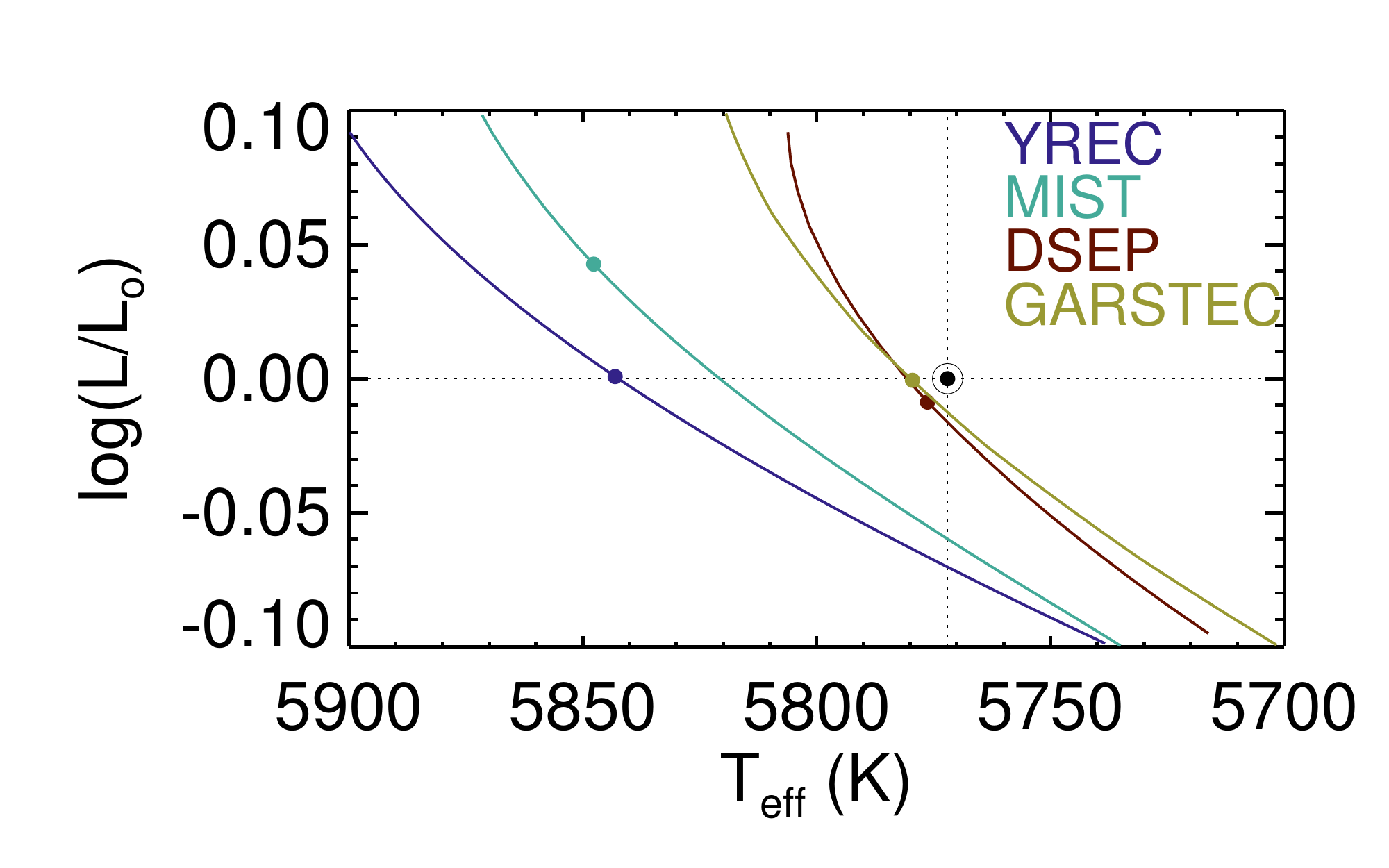}}
\caption{Different models make different predictions. Top: Predicted evolutionary tracks for solar metallicity stars of different masses for four different model grids. Bottom: Predictions for a solar mass, solar metallicity star. For each grid, the age of the sun \citep[4.57 Gyrs,][]{BahcallPinsonneault1995} is highlighted for comparison to the IAU solar values \citep[5772 K,][]{IAUsolar}, represented by the solar symbol.}


\label{Fig:modeldiffs}
\end{center}
\end{figure}

\subsection{Interpolation}
 We use the Python package \kiauhoku\ \citep{Claytor2020} to estimate the mass and age of stars given their metallicity, luminosity, and effective temperature for each grid of models. The packaged models grids for use with \kiauhoku\ are available on Zenodo under an open-source Creative Commons Attribution license\footnote{\dataset[10.5281/zenodo.4287717]{https://doi.org/10.5281/zenodo.4287717}} \citep{modelgrids}. For our purposes, we interpolate to find the best fit mass and age at one hundred effective temperatures between 4000 and 8000 K, one hundred luminosities between log(L/L$_\sun$)= -1 and 1.5, and four different metallicities ([Fe/H]=-1.0, -0.5, 0.0, and .5). 
 
 The \kiauhoku\ package works by resampling evolution tracks to equivalent evolutionary phases \citep[EEPs,][]{Dotter2016} and then interpolating stellar parameters given initial metallicity, initial mass, and EEP \citep[see][for more details]{Claytor2020}. While \kiauhoku\ does have a built in MCMC method to estimate parameters, a full MCMC run is extremely expensive given the number of grid points we wish to fit, and unnecessary for our purposes. Instead, we run the MCMC routine with 100 walkers with random starting masses between 0.6 and 2.0 \msun\ and equivalent evolutionary phases between the beginning of the main sequence (EEP 202) and the red giant branch bump (EEP 606). For each of these walkers, we record the sum squared difference between the search temperature and luminosity and the reported value. Each walker is run for only 10 steps, which does not allow the routine to converge. However, this initial run is enough to identify a mass and equivalent evolutionary point that gives a value close to the requested temperature and luminosity. We then use that initial guess in a minimization routine (scipy.optimize.minimize) and find that in the vast majority of cases, this is sufficient to estimate a mass and age 
 assuming models exist in the region of interest.

\begin{figure*}[tb]
\begin{minipage}{1.0\textwidth}
\begin{center}

\subfigure{\includegraphics[width=0.45\textwidth,clip=true, trim=0.5in 0in 0in 0in]{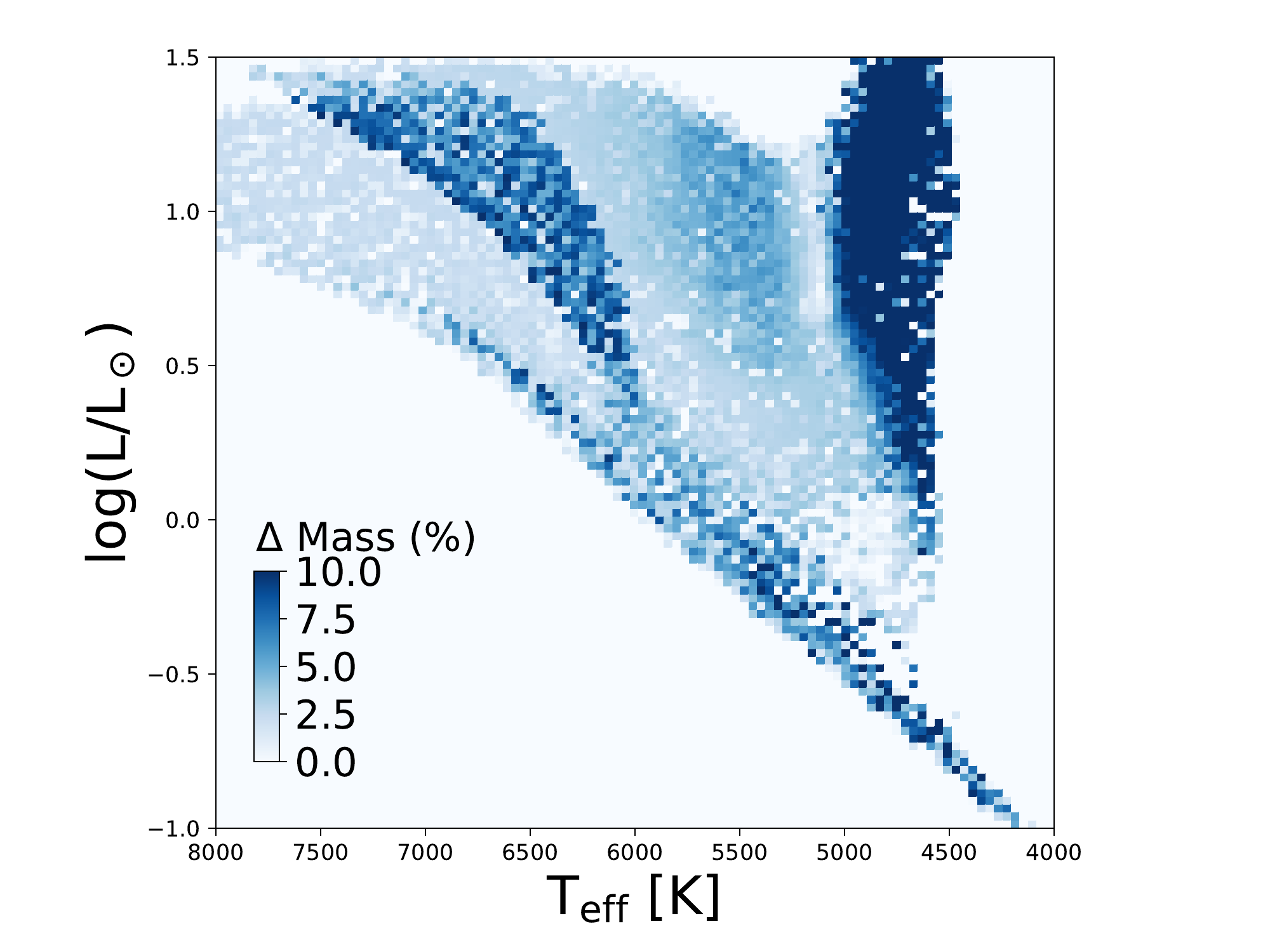}}
\subfigure{\includegraphics[width=0.45\textwidth,clip=true, trim=0.5in 0in 0in 0in]{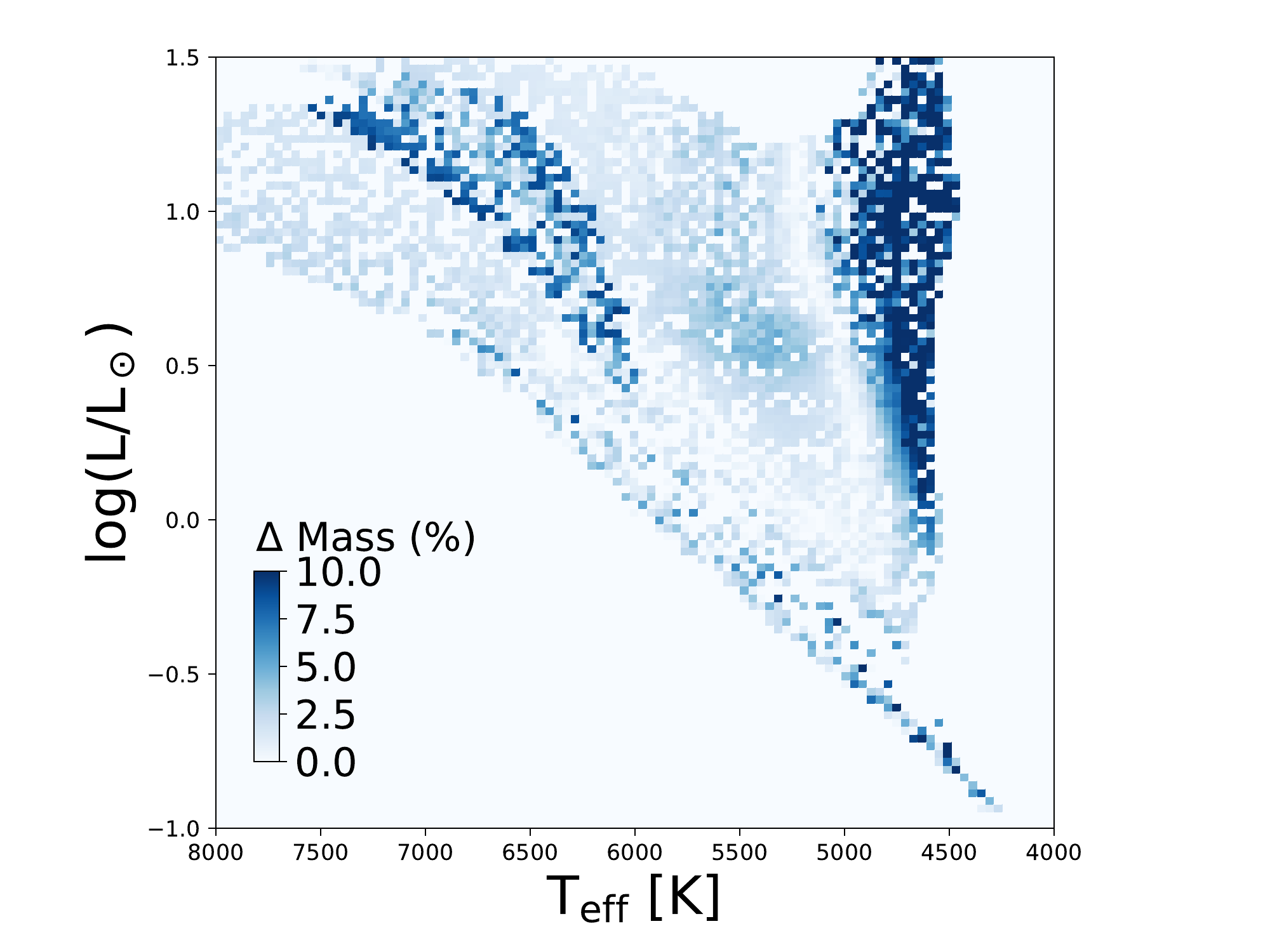}}
\subfigure{\includegraphics[width=0.45\textwidth,clip=true, trim=0.5in 0in 0in 0in]{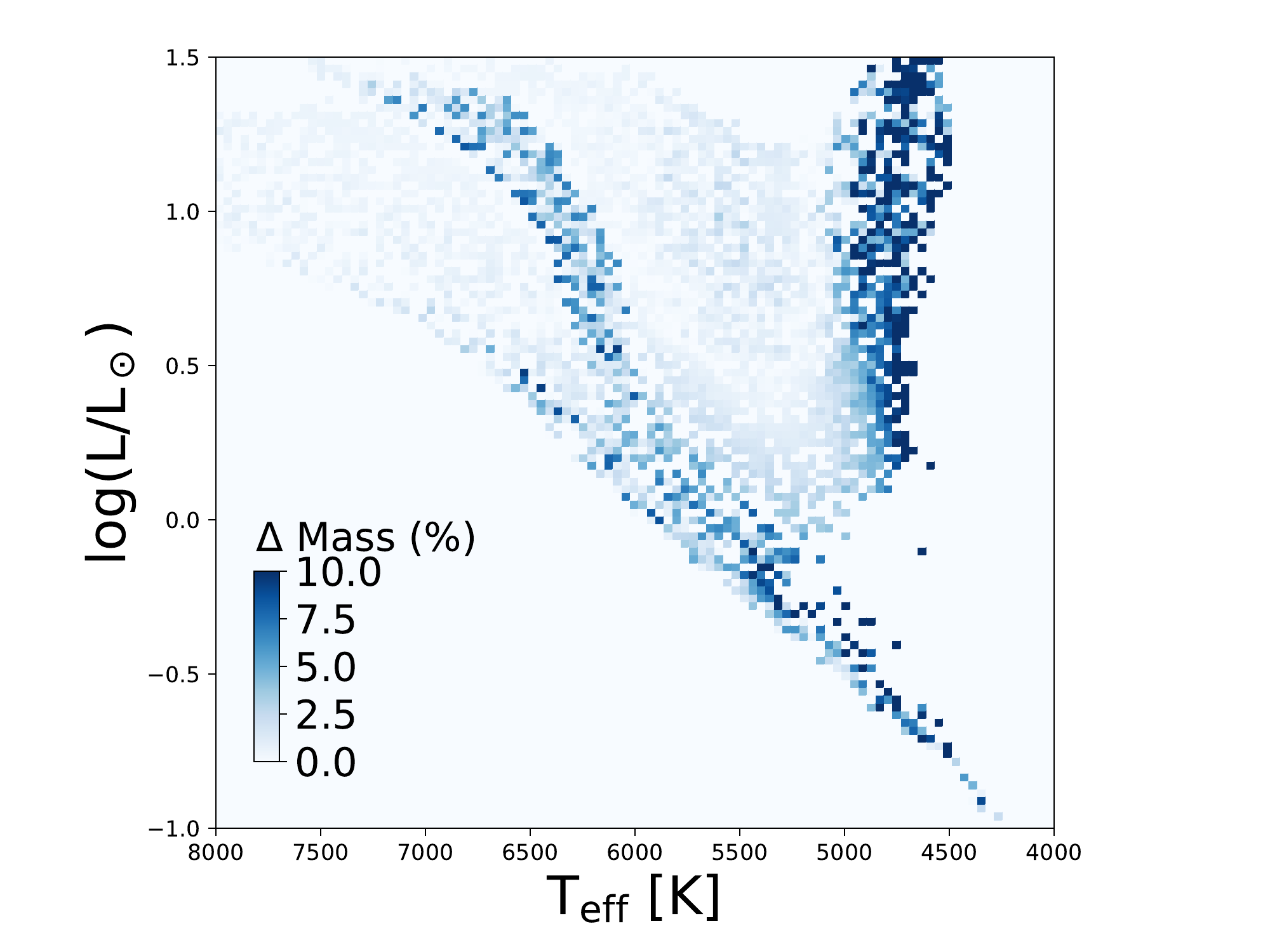}}
\subfigure{\includegraphics[width=0.45\textwidth,clip=true, trim=0.5in 0in 0in 0in]{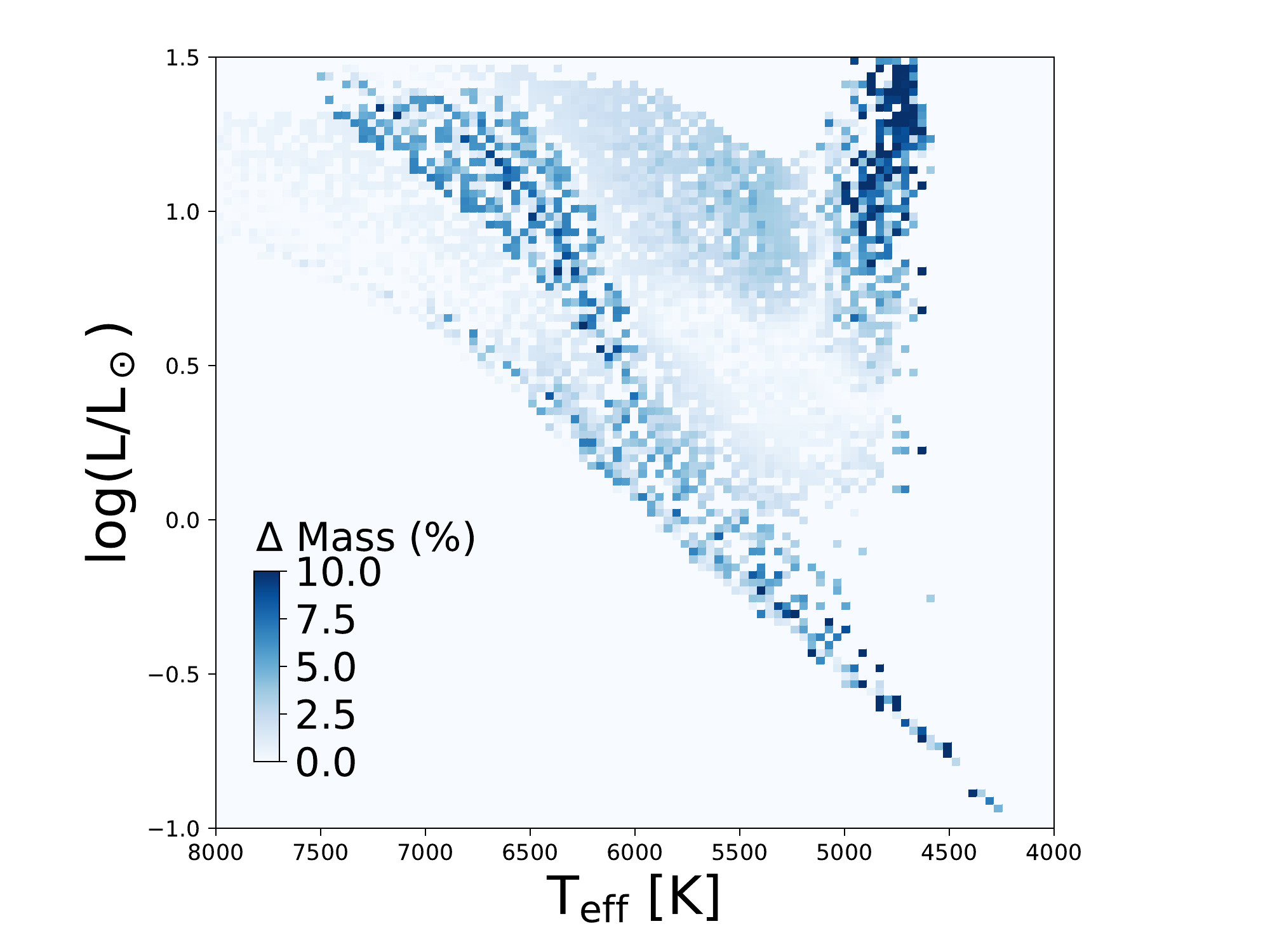}}
\caption{Maximal fractional offset in mass between model grids for stars at solar metallicity stars (top left: maximal difference, top right: MIST-YREC differences, bottom left: DSEP-YREC differences, bottom right: GARSTEC-YREC differences) as a function of temperature and luminosity. Offsets are largest in on the giant branch, and in the blue hook where overshoot choices are very important. }
\label{Fig:pairwise}
\end{center}
\end{minipage}
\end{figure*}

\begin{figure*}[tb]
\begin{minipage}{1.0\textwidth}
\begin{center}

\def\stackalignment{l}
\subfigure{\includegraphics[width=0.45\textwidth,clip=true, trim=0.5in 0in 0in 0in]{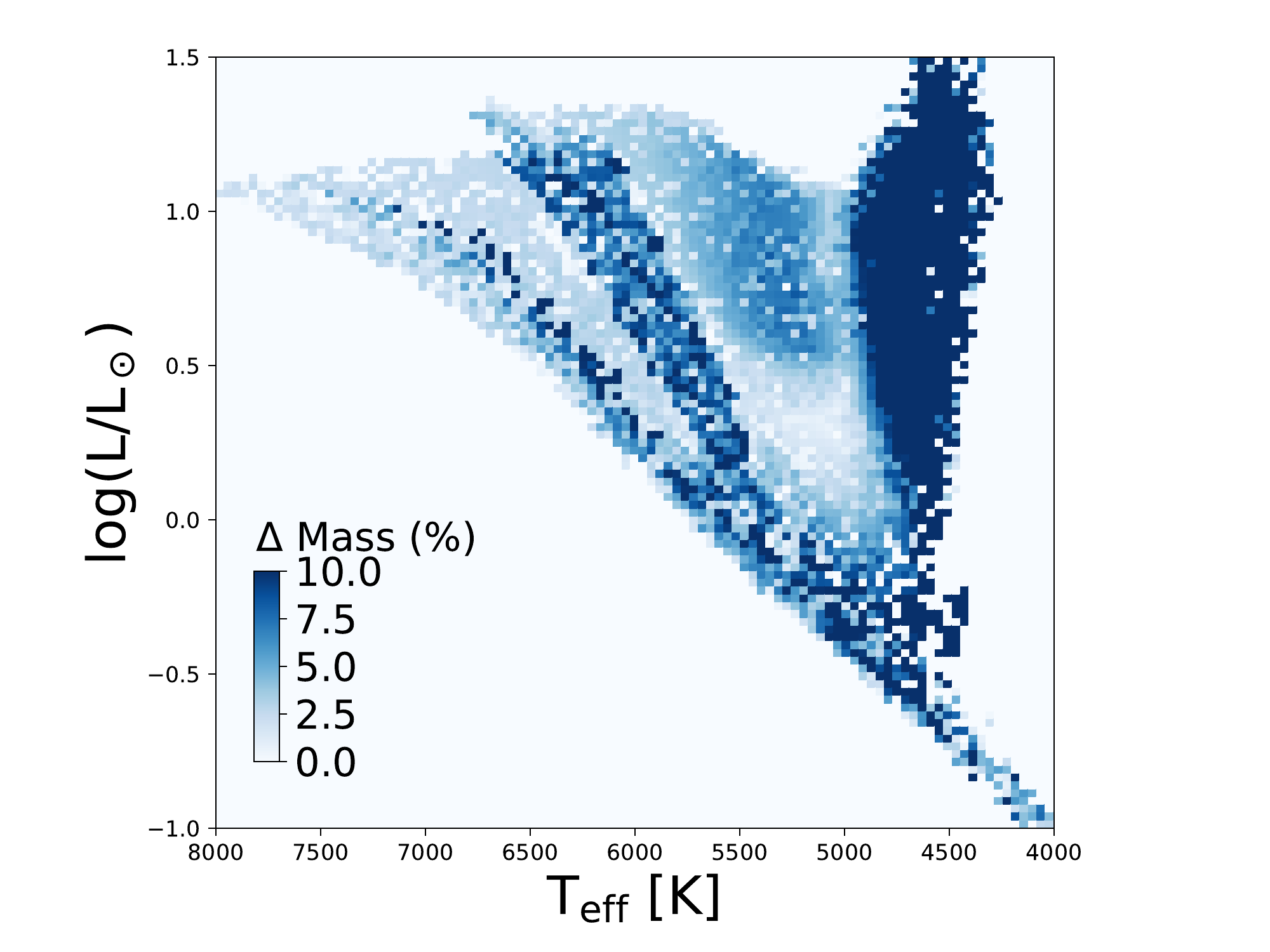}}
\subfigure{\bottominset{\includegraphics[width=0.13\textwidth,clip=true, trim=0.5in 0in 1in 0in]{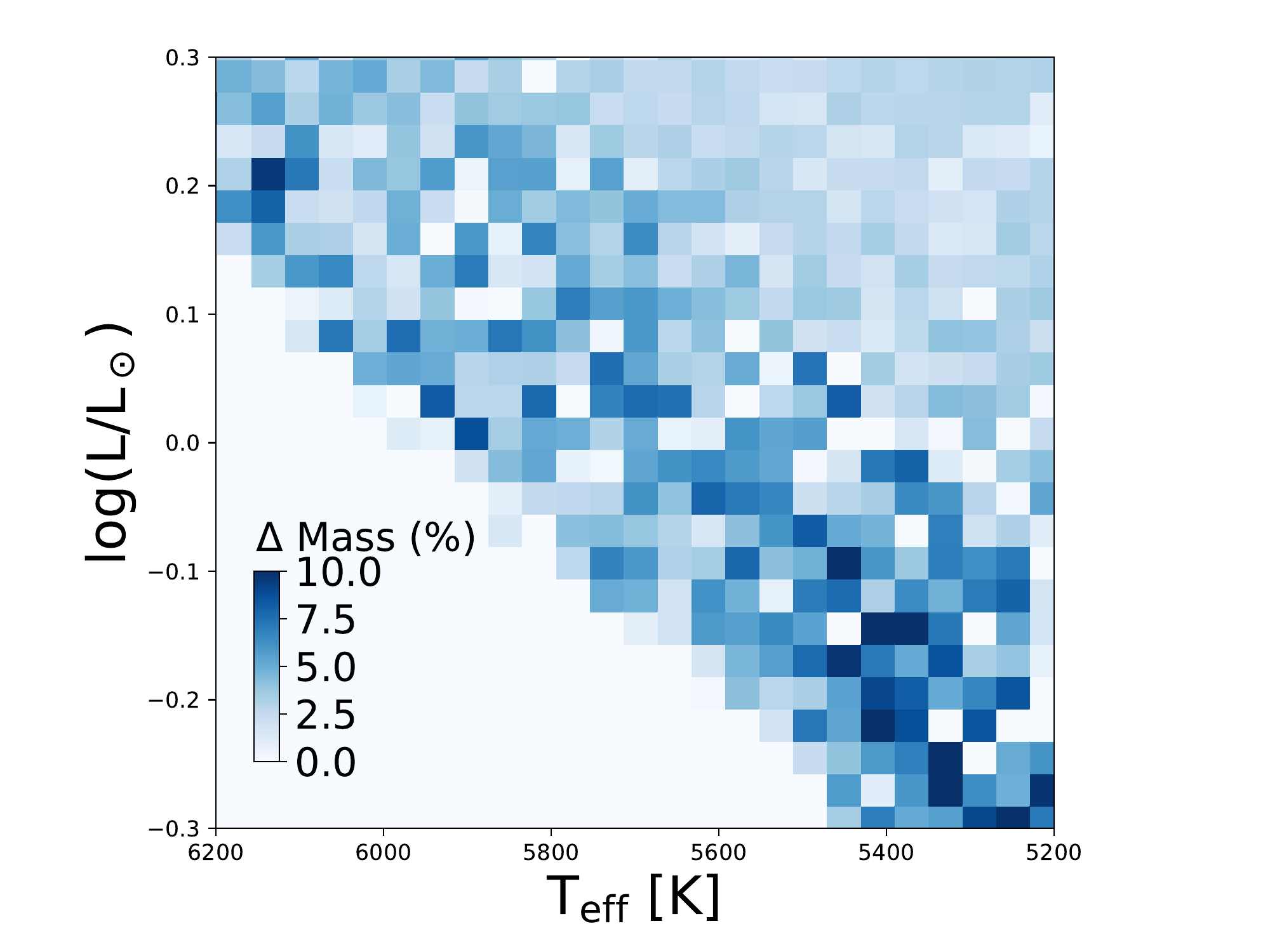}}{\includegraphics[width=0.45\textwidth,clip=true, trim=0.5in 0in 0in 0in]{./Figs/MassdiffplotmaxMdiff000000-eps-converted-to.pdf}}{30pt}{28pt}}
\subfigure{\includegraphics[width=0.45\textwidth,clip=true, trim=0.5in 0in 0in 0in]{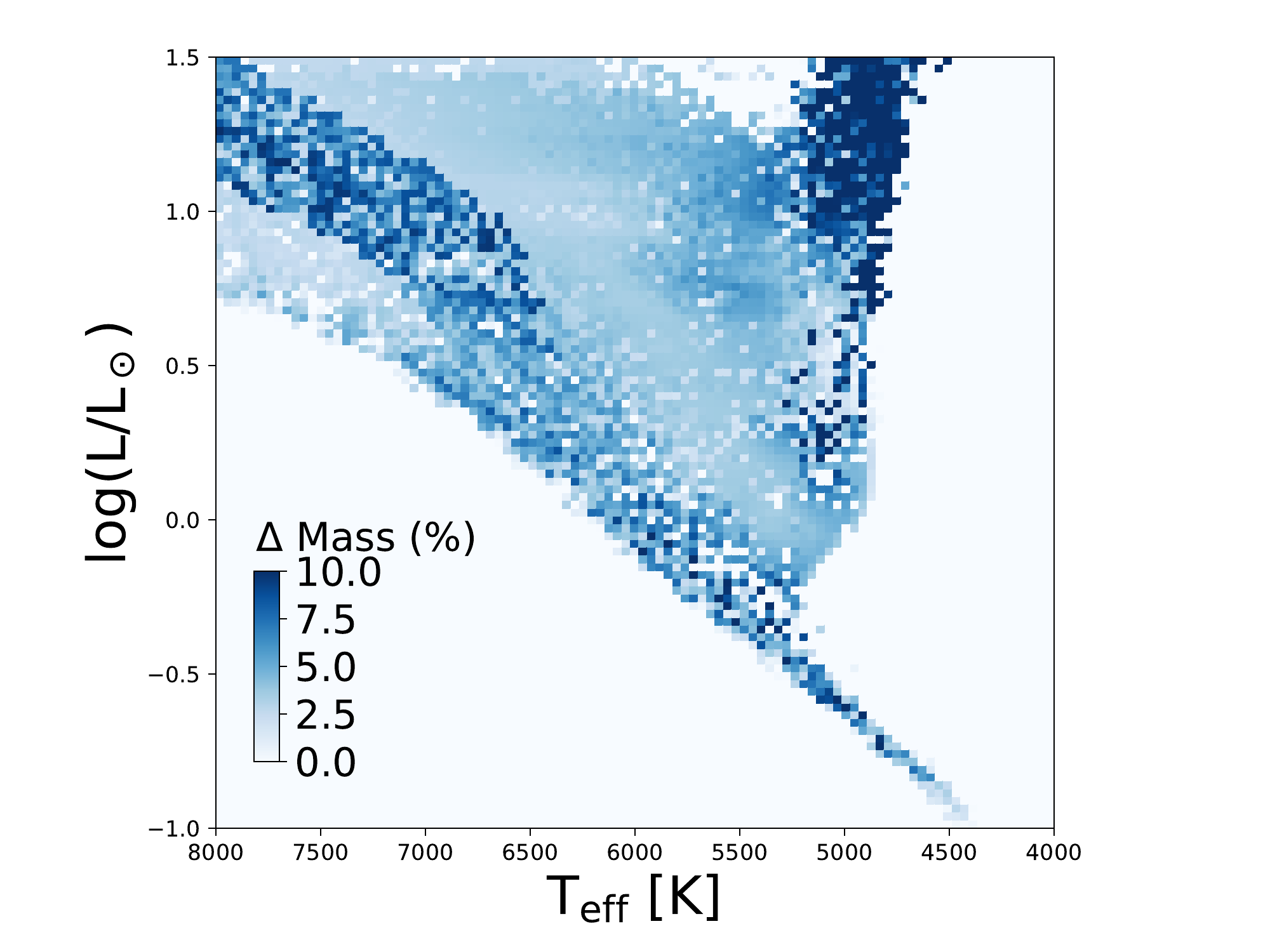}}
\subfigure{\includegraphics[width=0.45\textwidth,clip=true, trim=0.5in 0in 0in 0in]{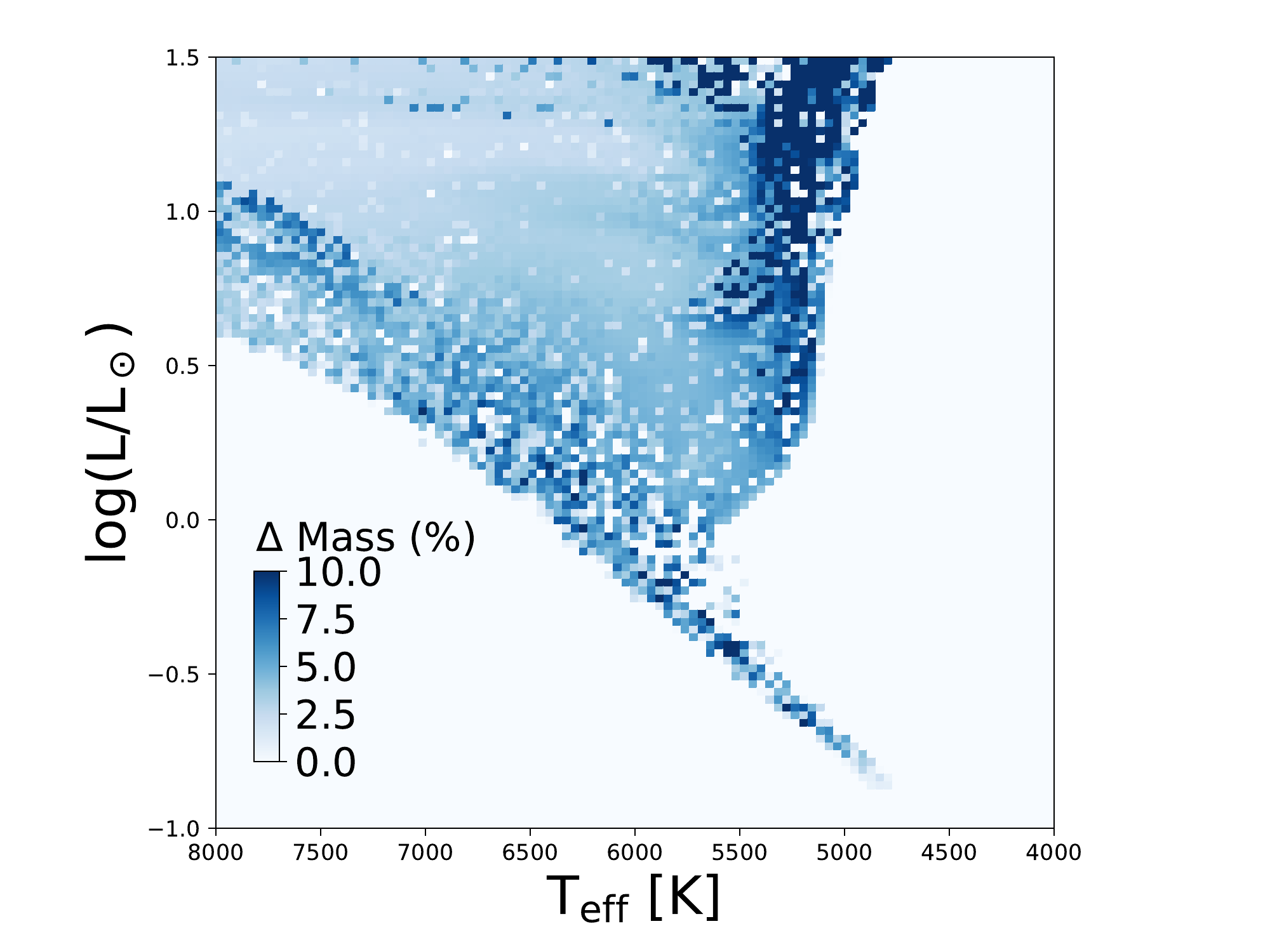}}
\caption{Maximal fractional offset in mass between model grids for stars at different metallicities (top left: +0.5, top right: solar, bottom left: -0.5, bottom right: -1.0) as a function of temperature and luminosity. Offsets are largest (darker colors) for stars near the giant branch, and $\sim$ 5 percent for most dwarfs and subgiants. }
\label{Fig:masses}
\end{center}
\end{minipage}
\end{figure*}

\subsection{Current Offsets}

As shown in Figure \ref{Fig:modeldiffs}, the different models make different predictions for the evolution of solar metallicity stars as a function of stellar mass. Zooming in on the solar mass models, it is evident that different calibration choices for some models lead to predictions that do not match the IAU values for the sun (5772 K) at the solar age (4.57 Gyrs). This does not necessarily imply that something is wrong with the generation of the model grid. There are well documented incompatibilities between recent spectroscopic solar abundance estimates \citep{Asplund2009} and helioseismic results \citep[e.g][]{Turck-Chieze2004,Buldgen2019}, and so modelers must either use an older abundance scale, or accept inconsistencies with solar parameters. In other cases, the physics of the models may have been optimized for stars in other parts of the HR diagram, or a slightly different solar temperature or age might have been used in the calibration.

We show in Figure \ref{Fig:pairwise} that it is not only the solar case but also across the HR diagram that the physics choices made lead to slight offsets in the locations of the model tracks. Models which use step overshoot versus exponential overshoot, for example, can have blue hooks with slightly different locations and shapes, which causes offsets in the masses inferred in that region. Similarly the exact location of the giant branch can be changed quite a bit by a number of physical assumptions \citep[e.g][]{Tayar2017, Choi2018a}, causing significant offsets in evolved stars. Looking at these plots, it is also clear that there is no single grid that is uniquely offset from the rest. Each pairwise set of models has places where they are more similar or more different, and it is the ensemble that likely indicates the true uncertainty in stellar evolution. 

For this reason, we argue that the maximal difference between grids of models should be considered an additional systematic error source for most applications. For example, many users want to estimate a stellar mass and age for a star given its measured properties, such as an inferred luminosity from \gaia\,  and an estimated temperature and metallicity from spectroscopy or photometry.
We show in Figure \ref{Fig:masses} that the maximal differences between models are a complicated function of luminosity, temperature, and metallicity. This means that even in the case of perfect measurements without uncertainties, the resulting estimated mass will depend on the model grid chosen. Specifically, we plot the maximum difference between any two grids of models at each point of interpolation and show that while offsets on the main sequence and subgiant branches are usually $\sim$5 percent in mass, systematic differences between models can be greater than 10 percent, particularly near the base of the giant branch. We also note that the estimated masses disagree more significantly in regions where the choice of overshoot is relevant. Around the blue hook, for example, the doubling back of the tracks as well as the choices of the amount and type of overshoot can impact the inferred mass by up to 7 percent, and for stars around 1.1 \msun, where a small convective core has developed, similarly large offsets can exist. 

\begin{figure*}[!htb]
\begin{minipage}{1.0\textwidth}
\begin{center}

\def\stackalignment{l}
\subfigure{\includegraphics[width=0.45\textwidth,clip=true, trim=0.5in 0in 0in 0in]{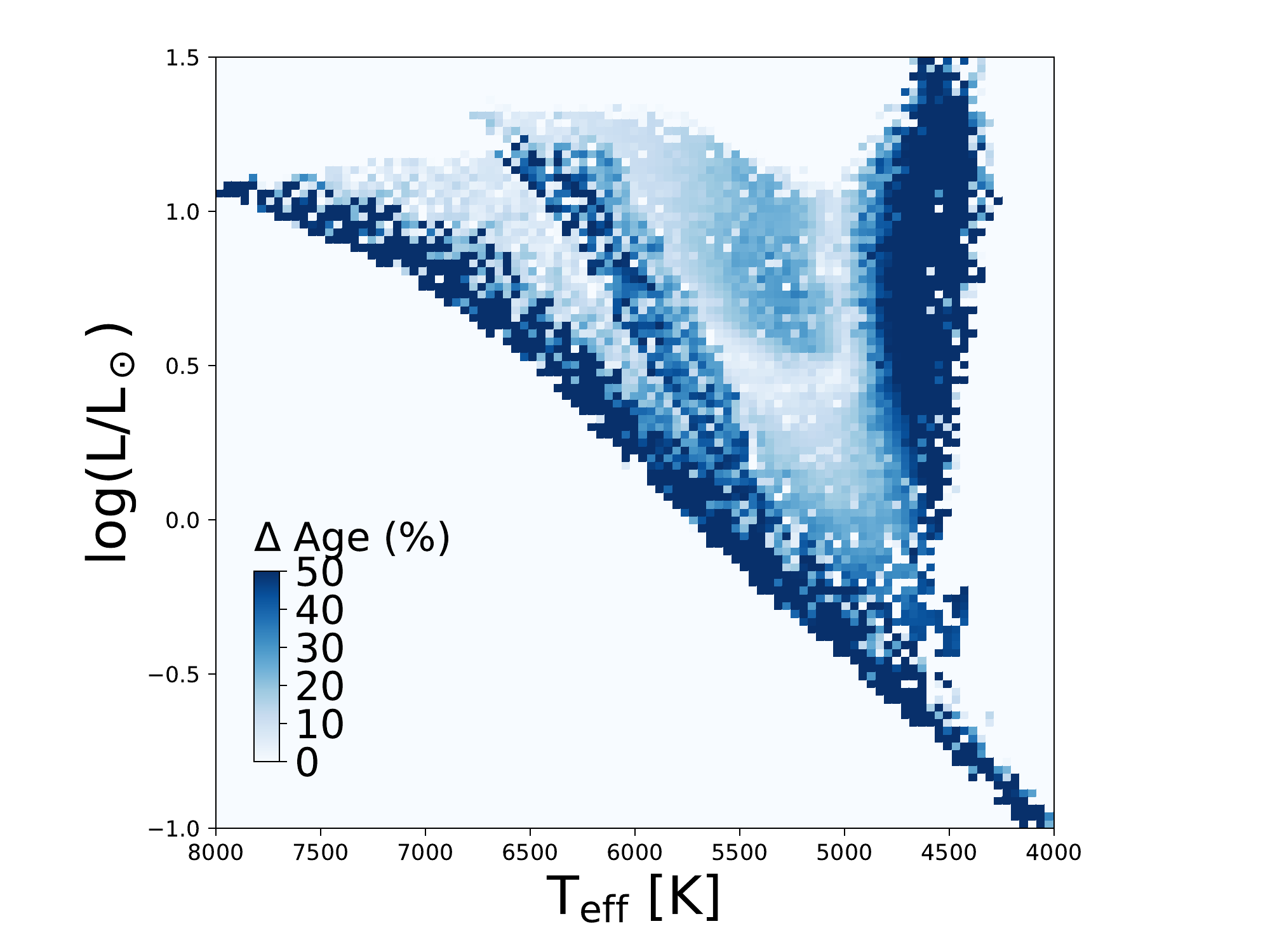}}
\subfigure{\bottominset{\includegraphics[width=0.13\textwidth,clip=true, trim=0.5in 0in 1in 0in]{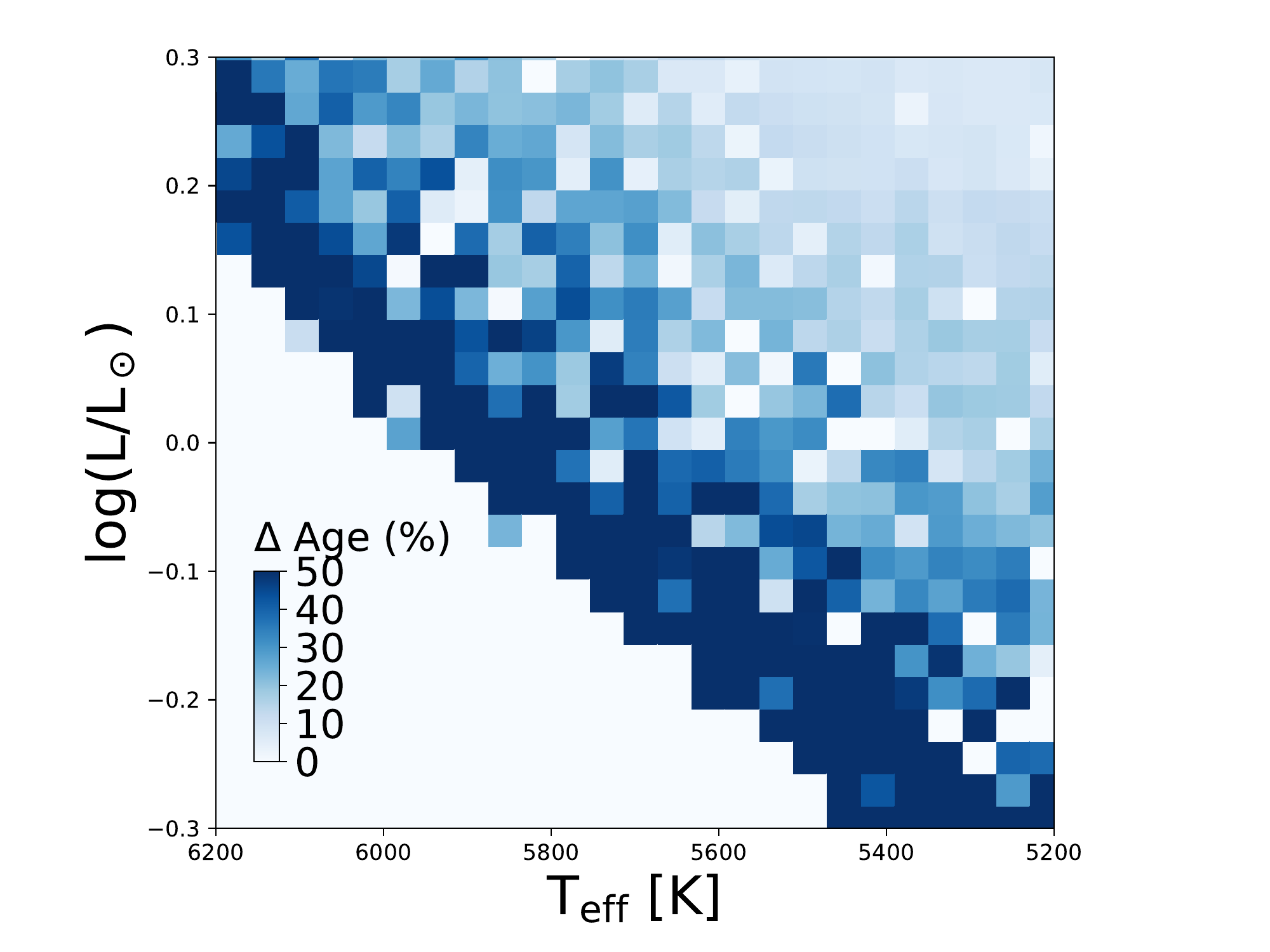}}{\includegraphics[width=0.45\textwidth,clip=true, trim=0.5in 0in 0in 0in]{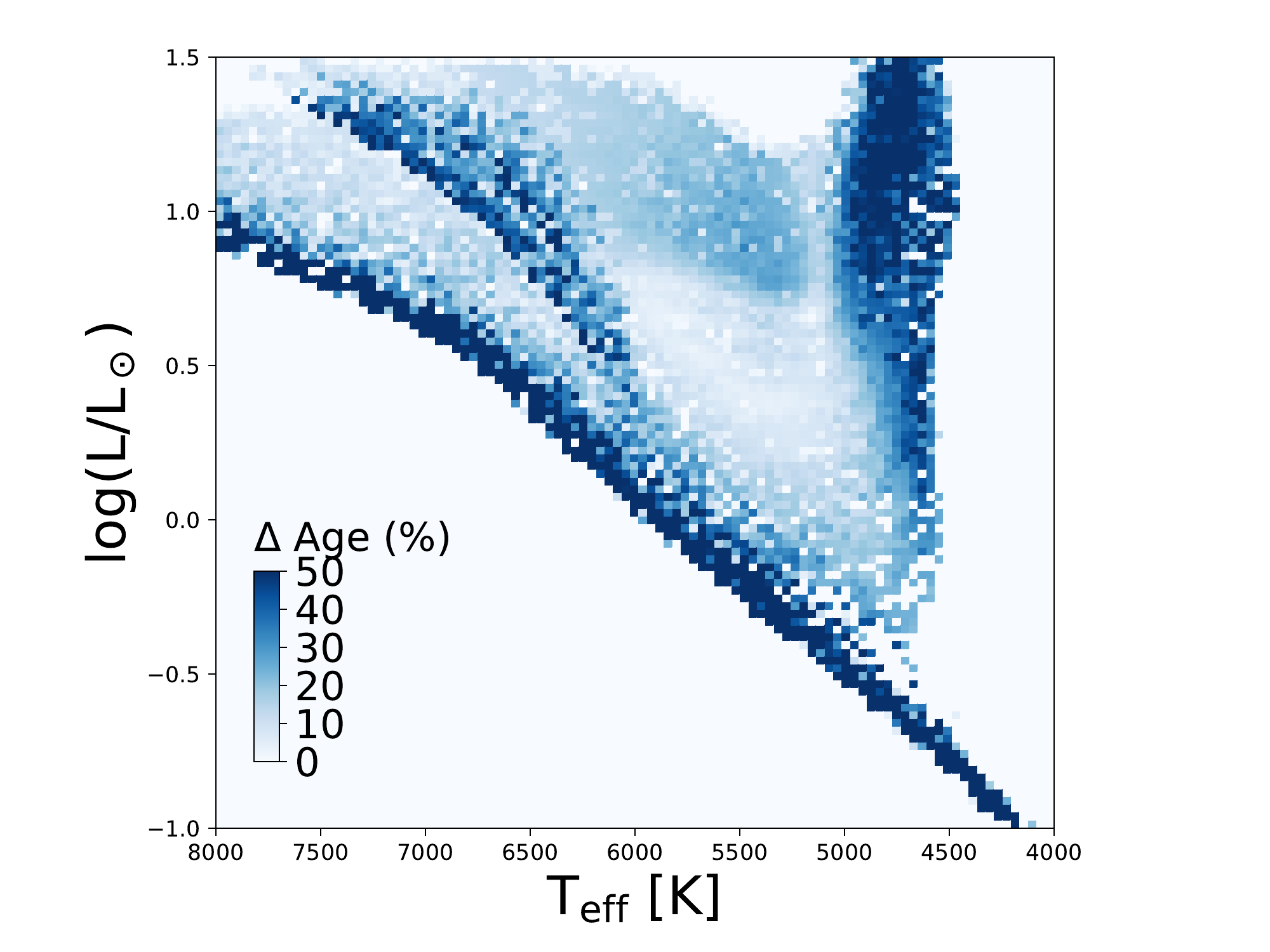}}{30pt}{28pt}}
\subfigure{\includegraphics[width=0.45\textwidth,clip=true, trim=0.5in 0in 0in 0in]{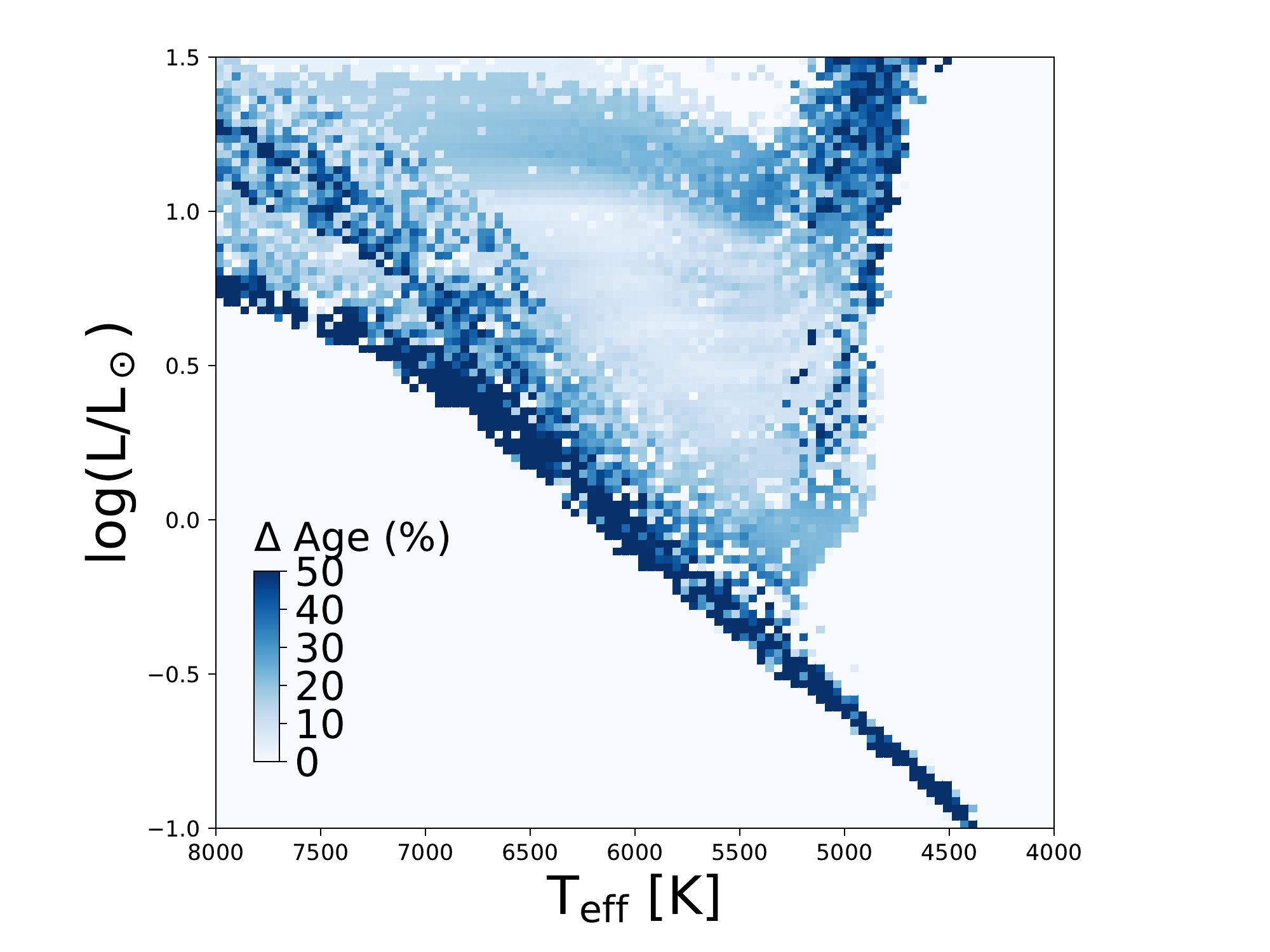}}
\subfigure{\includegraphics[width=0.45\textwidth,clip=true, trim=0.5in 0in 0in 0in]{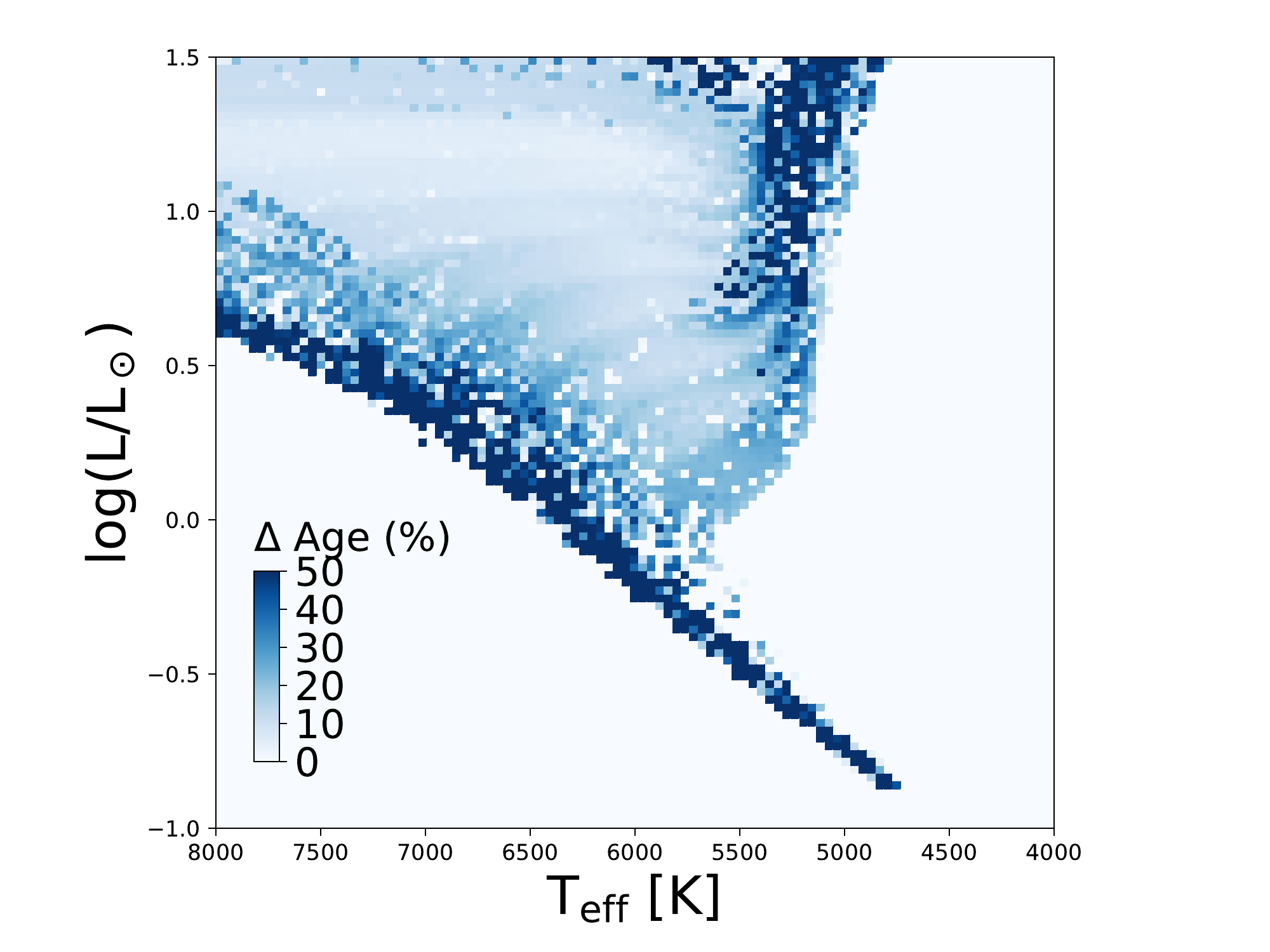}}

\caption{Maximal fractional offset in age between model grids for stars at different metallicities (top left: +0.5, top right: solar, bottom left: -0.5, bottom right: -1.0) as a function of temperature and luminosity. Offsets are largest (brighter colors) for stars near the zero age main sequence, and closer to $\sim$ 10 percent for most subgiant stars. }
\label{Fig:ages}
\end{center}
\end{minipage}
\end{figure*}

While the absolute errors in age follow a similar pattern, we show in Figure \ref{Fig:ages} that this is not the case for the fractional age uncertainties. When calculating ages, the difference between a zero age main sequence star in a model grid that tends to be hotter and a main sequence turnoff star in a model grid that tends to be cooler are very small in mass, but almost one hundred percent in age. This means that the dominant age uncertainty for some stars may be the choice of model grid and this uncertainty should not be ignored, particularly near the zero age main sequence.

\subsection{Mitigating Offsets}

\subsubsection{\kiauhoku\ Tools}
{The discrepancies between stellar model grids demand an accounting of systematic uncertainties from grid to grid. To aid future efforts in this, we provide a web-based Google Colab interface to \kiauhoku\footnote{\url{https://colab.research.google.com/drive/1tmOHsat2g2-fYU8Y2UJIdV6VWCI9RjTf?usp=sharing}}.
This tool allows a user to input observables for an individual star and compute model mass and age offsets in the four currently implemented model grids without downloading anything. While this makes computing systematic uncertainties for individual stars more convenient, for larger datasets we recommend installing \kiauhoku\footnote{Install from the Python Package Index by typing ``\texttt{pip install kiauhoku}" into the command line.} and using the Python interface.}

\subsubsection{Improving Stellar Models}

While our previously described tool provides a method for accounting for uncertainties on stellar parameters due to uncertainties in stellar models, this is obviously not the optimal solution. Removing all uncertainties on stellar physics also seems unlikely to occur in the near future, especially since three dimensional simulations of stellar interiors over cosmic time are still computationally unfeasible. However, we suggest that in the near future, it will be possible to at least calibrate the models that we do have for many interesting regions of parameter space. Asteroseismology, in particular, is allowing the estimation of the masses of tens of thousands of stars, which, when combined with spectroscopic characterization, can be used to check the calibration of stellar models \citep[e.g.][]{Tayar2017}. Work in open clusters allows the checking of models as a function of age \citep[e.g.][]{Choi2018b, Sandquist2020}. Finally, the careful characterization of double lined eclipsing binaries \citep{ClaretTorres2019}, or in some cases even single lined eclipsing binaries \citep{Stevens2018} can constrain mass and age simultaneously. While the accuracy and precision of all of these measurements are still being refined, \citep[see e.g][]{Gaulme2016}, they represent an exciting opportunity to substantially improve our ability to estimate the masses and ages of stars to high precision. Combining these samples to select a set of $\sim$100 of the best characterized stars at a range of metallicities, temperatures, and luminosities could provide a set of benchmarks for modelers to validate new stellar evolution grids, analogous to unit tests in computer science, the \gaia\ benchmark stars in spectroscopy, or the way in which 16 Cyg has functioned for solar-like asteroseismologists.   




\section{Worked Examples}
\subsection{$\pi$ Mensae }
$\pi$ Mensae was the host of the first planet discovered by \tess\ \citep{huang18, gandolfi18}. 
Both discovery papers adopt effective temperatures with uncertainties $<$\,1\%, which individually differ by $\sim$\,3\,$\sigma$. They also adopt stellar radii and masses with uncertainties of $1-2$\% and $3-4$\%, respectively.


To estimate the expected systematic errors from models only (ignoring differences in effective temperatures), we adopt the derived properties from \citet{huang18}: \teff= 6037 $\pm$ 45 K, L$_{\star}$=1.444 $\pm$ 0.02 L$_\sun$, [Fe/H]= 0.08 $\pm$ 0.03. Applying these values to the method described in Section 3 yields  grid masses of 1.090 \msun, 1.099 \msun, 1.110 \msun, and 1.123 \msun\  for YREC, MIST, DSEP, and GARSTEC respectively. All of these are consistent with the recently derived asteroseismic scaling relation mass from TESS 20-second cadence observations in Sectors 27 and 28 (1.145 $\pm$ 0.08, D. Huber et al, in prep) but only some of these estimates are consistent with the 1.09 $\pm$ 0.03 quoted by \citet{huang18}, suggesting that in this case, the systematic errors are comparable to or dominate the random observational uncertainties. We also note that the range of model ages (3.50, 2.28, 2.33, and 1.73 Gyrs for YREC, MIST, DSEP, and GARSTEC respectively) span a similar range to the observational uncertainties, quoted as 2.98 $_{-1.3}^{+1.4}$ Gyrs. Thus, the systematic uncertainties should not be neglected in this regime, and should be added in quadrature to the observational uncertainties.

\subsection{TOI-197}
TOI-197 was the first example of an oscillating planet hosting star with \tess\ \citep{Huber2019}. The host star is near the end of the subgiant phase and has a well constrained temperature (5080 $\pm$ 90 K), metallicity ([Fe/H]=-0.08 $\pm$ 0.08), and luminosity (5.15 L$_\sun$ $\pm$ 0.17). Given these parameters the different model grids would have inferred masses of 1.252 \msun, 1.210 \msun, 1.137 \msun, and 1.194 \msun\ for YREC, MIST, DSEP, and GARSTEC respectively, most of which are consistent with the quoted asteroseismic mass (1.212 $\pm$ 0.074 \msun). 

 This star illustrates the challenges of estimating masses from stellar models as stars approach the red giant branch. Since the models are not substantially separated in temperature, precise mass estimates are impossible without exceptionally good observations or additional information. Given the full range of observational uncertainties, the MIST grid of models would have allowed for masses between 1.04 and 1.33 \msun. However, it must be noted that the ten percent systematic uncertainty between model grids is not entirely negligible even in this challenging observational regime. The range of ages in this regime is also significant, with models giving ages of 4.9, 4.9, 6.6, and 5.6 Gyrs for YREC, MIST, DSEP, and GARSTEC respectively. This is a $\sim$30 percent systematic spread, which needs to be added in quadrature to the random uncertainty.


\section{Conclusions}
The recent advent of high-precision astrometry from \gaia\ and large spectroscopic surveys have enabled precise measurements of single field star properties 
such as radius, mass, and age, which in principle allow exciting new explorations into the demographics, compositions, and atmospheres of planets that orbit these stars. However, we have demonstrated here that caution is required when quoting very precise fundamental properties for stars and exoplanets, as systematic uncertainties can dominate the error budget for stellar properties. For example, the uncertainty on the fundamental temperature scale from interferometry, in combination with the uncertainties on flux scales, extinctions, and bolometric corrections, in most cases limits temperature estimates to $\sim$\,2\%, luminosities to $\sim$\,2\%, and radii to $\sim$\,4\%, a factor of 2--4 higher than the typically quoted uncertainties in the recent literature. 

Estimating stellar masses from stellar models also has significant uncertainty, 
as different grids of stellar models will disagree on the inferred mass and age at the few percent level, due to uncertainties on the physics of the stellar interior. We have shown that these offsets between models are luminosity, temperature, and metallicity dependent, and are commonly on the order of $\sim$5\% in mass and $\sim$20\% in age, although they can be substantially larger. This uncertainty from the model choices can be as large as or larger than the uncertainties from observation in some cases, and as such should be considered in analyses. Most properly, this should be done by perturbing all of the uncertain physics, and comparing to external checks to properly calibrate the models in the regions of interest. In practice, the precision and volume of data necessary to undertake such a study is only now becoming available. In the interim, we recommend using the range of results returned from various available model grids as a measure of the systematic uncertainty of a star's mass and age, and we have provided open-source software to estimate these values given the stellar parameters so that they can be added in quadrature to the observational uncertainties. 

We note that the uncertainty estimates we provide here are a guideline for typical stars, and that it may be possible to do better in carefully studied individual cases or in stars with additional constraints. 
Future improvements in stellar model physics as well as a larger number of dedicated fundamental measurements through interferometry and space-based spectrophotometry will be required to reduce systematic errors in host star (and thus exoplanet) properties to the level of precision that current observational datasets enable. However, such careful work has the potential to change what we can discover about stars and their planets and is thus a worthwhile effort. 

\begin{acknowledgements}
We thank Aaron Dotter and Aldo Serenelli for their assistance in using their grids, and Gail Schaefer for help with compiling angular diameter measurements from the literature.
We also acknowledge helpful discussions with Andrew Mann, Marc Pinsonneault, Jim Davenport, Ellie Abrahams, Gail Schaefer and the online.tess.science eclipsing binary working group. 
 J.T. acknowledges that support for this work was provided by NASA through the NASA Hubble Fellowship grant No.51424 awarded by the Space Telescope Science Institute, which is operated by the Association of Universities for Research in Astronomy, Inc., for NASA, under contract NAS5-26555 and by
NASA Award 80NSSC20K0056. The authors thank the Kavli Institute for Theoretical Physics for hosting the Exostar19 meeting and acknowledge that this research was supported in part by the National Science Foundation under Grant No. NSF PHY-1748958. 
Z.R.C. acknowledges support from the TESS Guest Investigator Program (80NSSC18K18584) and the Heising Simons Foundation.
D.H. acknowledges support from the Alfred P. Sloan Foundation and the National Aeronautics and Space Administration (80NSSC19K0597), and the National Science Foundation (AST-1717000).


\end{acknowledgements}

\bibliographystyle{apj} 
\bibliography{ms,references}
\end{document}